\documentclass{article}
\usepackage{spconf,amsmath}
\usepackage[pdftex]{graphicx}
\usepackage{multirow}
\usepackage{here}
\usepackage{multicol}
\newcommand{\vect}[1]{{\mbox{\boldmath $#1$}}}
\newcommand{\bhline}[1]{\noalign{\hrule height #1}}  

\title{Flow-based self-supervised density estimation for anomalous sound detection}
\name{Kota Dohi, Takashi Endo, Harsh Purohit, Ryo Tanabe, Yohei Kawaguchi}
\address{Research and Development Group, Hitachi, Ltd.\\
1-280, Higashi-koigakubo, Kokubunji-shi, Tokyo 185-8601, Japan\\
\texttt{kota.dohi.gr@hitachi.com}}

\begin{document}
\ninept
\maketitle
\begin{abstract}
  To develop a machine sound monitoring system, a method for detecting 
  anomalous sound is proposed. 
  Exact likelihood estimation using Normalizing Flows is a promising technique for unsupervised anomaly 
  detection, but it can fail at out-of-distribution detection since the likelihood is affected by the smoothness of the data.
  To improve the detection performance, we train the model to assign higher likelihood 
  to target machine sounds and lower likelihood to sounds from other machines of the same machine type.
  We demonstrate that this enables the model to incorporate a self-supervised c\-l\-a\-ssification-based approach. 
  Experiments conducted using the DCASE 2020 Challenge Task2 dataset showed that the proposed method improves the AUC by 4.6\% 
  on average when using Masked Autoregressive Flow (MAF) and by 5.8\% when using Glow, which is a significant 
  improvement over the previous method.
\end{abstract}

\section{Introduction}
\label{sec:intro}
As a number of companies worldwide are facing a shortage 
of skilled maintenance workers, the demand for an automatic sound-monitoring system has been increasing. 
Unsupervised anomaly detection methods are often adopted for this system 
since anomalous data can rarely be obtained \cite{Suefusa2020,Koizumi2020}.

For unsupervised anomaly detection, generative models such as Variational Auto Encoder (VAE), which use 
approximate likelihood and reconstruction error to calculate
anomaly scores, have been utilized.
Normalizing Flows (NF) \cite{Tabak2013,Dinh2015} is another promising generative model thanks to its 
ability to perform exact likelihood estimation.
However, this model fails at out-of-distribution detection since it assigns higher likelihood to 
smoother structured data \cite{Nalisnick2019,Choi2019,Serr2020}. 

Self-supervised classification-based approach 
is another way for detecting anomalies when
sound data from multiple machines of the same machine type is available \cite{Giri2020}. 
In self-supervised learning, a model is trained for a main task with another task called an auxiliary task
to improve the performance on the main task. In self-supervised classification-based approach, the auxiliary
task is to train a classifier that predicts a machine ID assigned to each machine. If the classifier misclassifies
the machine ID of a sound data, the sound data is regarded as anomalous. 
Although this approach improves the detection performance on average, it shows 
significantly low scores on some machines.

In this paper, we propose a self-supervised density estimation method using NF. 
Our method uses sound data from one machine ID to detect anomalies (target data) and sound data from 
other machines of the same machine type (outlier data), and the model is trained to assign higher likelihood to 
the target data and lower likelihood to the outlier data. 
This method is a self-supervised approach because it improves the detection performance on one machine ID 
by introducing an auxiliary task in which the model discriminates the sound data of that machine ID (target data) from 
sound data of other machine IDs with the same machine type (outlier data). 
Also, since the method increases the likelihood of the 
target data, this method can also be similar to the unsupervised approach.

We evaluated the detection performance of our method using six types of 
machine sound data. Glow \cite{Kingma2018} and MAF \cite{Papamakarios2018} were utilized as 
NF models.
Experimental results showed that our method outperformed unsupervised approaches 
while showing more stable detection performances compared to the self-supervised classification-based approach.

\section{Problem Statement}
\label{sec:statement}
Anomalous sound detection is a task to identify whether a machine
is normal or anomalous based on the anomaly score a trained model calculates from its sound.
Each input sound data is determined as anomalous data if the anomaly score of the data 
exceeds a threshold value.
We consider the unsupervised anomalous sound detection where only normal sound is 
available for training.
We assume sound data of multiple machines with the same machine type is 
available. This is a realistic assumption since multiple machines of the same type are often installed in factories.
This problem setting is the same as that in DCASE 2020 Challenge Task2 \cite{koizumi2020_2}.

\section{Relation to prior work}
\subsection{Improving the detection performance of NF}
Researchers have proposed various methods to modify the likelihood assigned by NF models.
Serra et al. \cite{Serr2020} used input complexity to modify the likelihood. 
However, a compression algorithm has to be chosen to calculate the input complexity.
Ren et al. \cite{Ren2019} used the likelihood ratio between the likelihood of the input data and the background component,
but parameters have to be tuned for modeling the background. 
Hendrycks et al. \cite{Hendrycks2019} used auxiliary datasets of outliers to improve the detection performance.
However, 
the proposed loss function can destabilize the detection performance.
Kirichenko et al. \cite{Kirichenko2020} proposed a loss function to distinguish the in-distribution data from 
out-of-distribution data by means of a supervised approach. 
The authors argued that
this method cannot be used to detect out-of-distribution data not included in the training dataset.
We show that this method can be used for detecting anomalies
if sound data of the same machine type as the target data is used for the outlier data in the training dataset.

\subsection{A self-supervised approach for anomaly detection}
In DCASE 2020 Challenge Task2, top rankers trained classifiers to
predict a machine ID of the data \cite{Giri2020, Primus2020}. 
This approach assumes that the classifier can output
a false machine ID if the data is anomalous. 
Giri et al. \cite{Giri2020} named this approach a self-supervised classification-based approach.
However, we found that this approach can fail on some machine IDs, and the detection performance degrades
significantly.
To stabilize the detection performance, our proposed method incorporates the unsupervised approach by
using the NLL to distinguish the target data from the outlier data.

\section{conventional approach}
\label{sec:conventional}
Normalizing Flows (NF) is a series of invertible transformations 
between an input data distribution $p(\vect{x})$ and 
a known distribution $p(\vect{z})$. 
Anomaly score can be calculated by the negative 
log likelihood (NLL) of the input data \cite{Schmidt2019, Yamaguchi2019, Ryzhikov2019, Dias2020}.
However, this score is dependent on the smoothness of the input data and fails at out-of-distribution detection.


Kirichenko \cite{Kirichenko2020} proposed a loss function to distinguish the in-distribution data 
from the out-of-distribution data by using a supervised approach:
\begin{equation}
  \begin{split}
  L           = &\frac{1}{N_{D}} \sum_{\vect{x}\in D} \textrm{NLL}(\vect{x}) \\
                &- \frac{1}{N_{OOD}} \sum_{\vect{x}\in OOD} \textrm{NLL}(\vect{x}) \cdot I[\textrm{NLL}(\vect{x})<c],
  \label{kiri_loss}
  \end{split}
\end{equation}
where $N_{D}$ is the number of the in-distribution data in each batch,  $N_{ODD}$ 
is the number of the out-of-distribution data in each batch that satisfies the condition in the indicator function
$I[\cdot]$, and $c$ is a threshold value. 

\section{Proposed Approach}
\label{sec:proposed}
To overcome the problems of NF models and the s\-e\-l\-f-\-s\-u\-p\-ervised classification-based approach, our method 
attempts to assign higher likelihood to the target data and lower likelihood to the outlier data using NF.

\begin{figure}[t]
  \centering
    \includegraphics[width=0.60\hsize,height=0.12\vsize,clip]{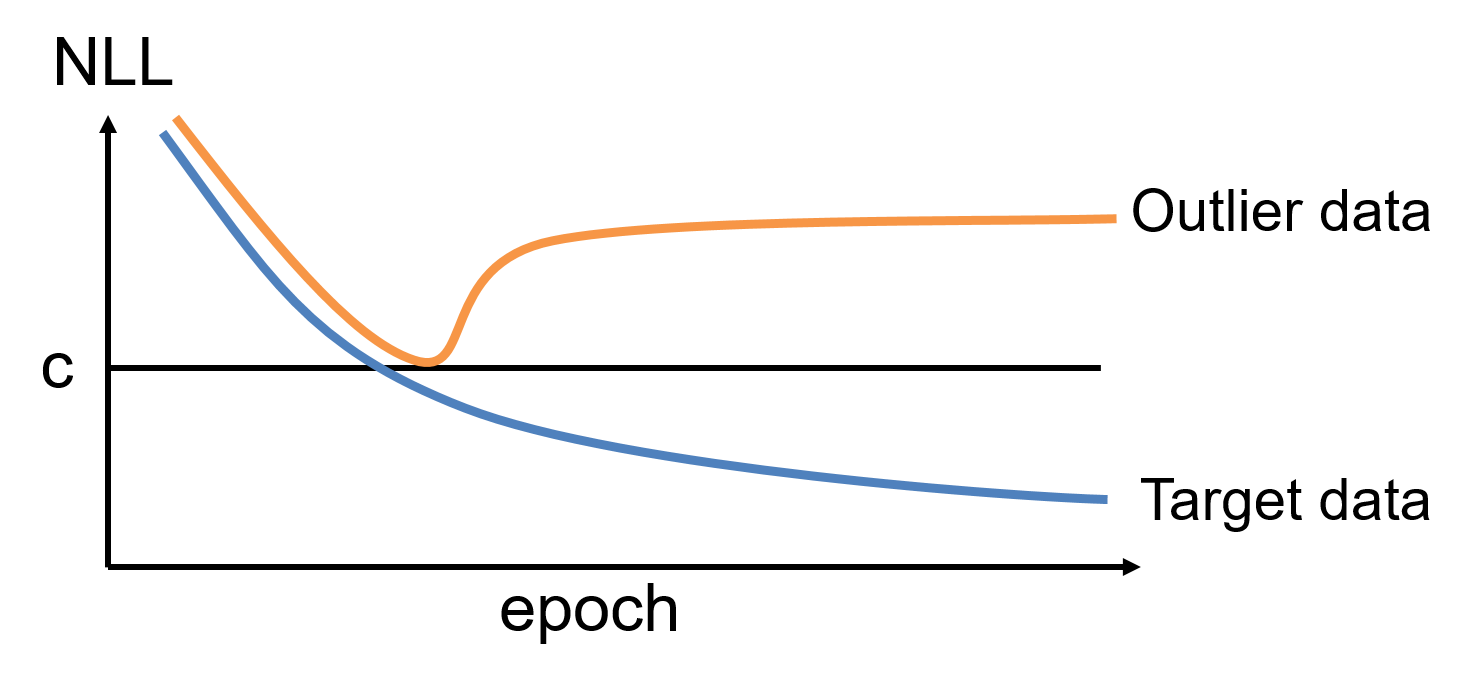}
    \caption{Illustration of the proposed approach.}
  \label{fig:idea}
\end{figure}

We train a model for each machine ID, where the data with that ID is the target data $\vect{x}_{t}$ 
and other machine sounds of the same machine type is the outlier data $\vect{x}_{e}$. 
Assume $\vect{x}_{t}$ and $\vect{x}_{e}$ consist of components specific to their machine IDs ($\vect{x}_{t0}$, 
$\vect{x}_{e0}$)
and components shared across the same machine type ($\vect{x}_{c}$). The likelihood of $\vect{x}_{t}$ and $\vect{x}_{e}$ 
can be written as
\begin{equation}
  p(\vect{x}_{t}) = p(\vect{x}_{t0})p(\vect{x}_{c}),
  \label{target_likelihood}
\end{equation}
\begin{equation}
  p(\vect{x}_{e}) = p(\vect{x}_{e0})p(\vect{x}_{c}).
  \label{outlier_likelihood}
\end{equation}
If the model is trained to assign higher likelihood to $\vect{x}_{t}$ and lower likelihood to $\vect{x}_{e}$, 
only components specific to the machine ID ($\vect{x}_{t0}$, 
$\vect{x}_{e0}$) will affect the likelihood. 
Therefore, we can improve the detection performance of a NF model by introducing an auxiliary task in which we train 
the NF model to discriminate the target data $\vect{x}_{t}$ from the outlier data $\vect{x}_{e}$ by the likelihood.
When testing the model, the NLL is used as the anomaly score.
The anomaly score will be higher when the data structure is different from the target data specific component 
$\vect{x}_{t0}$ or close to the outlier data $\vect{x}_{e0}$. 
The former case corresponds to anomaly detection by the unsupervised approach and the latter case to
the self-supervised approach.
Therefore, this idea can benefit from both the unsupervised and the self-supervised
classification-based approach. As illustrated in Fig.\ref{fig:idea},
the model is trained to decrease the NLL until the NLL of 
the outlier data reaches a threshold $c$. Then, the model is penalized so that the NLL of the outlier data
can be higher than the target data.
This idea can be realized by using the loss function in (\ref{kiri_loss}), as
\begin{equation}
  \begin{split}
  L = &\frac{1}{N_{target}} \sum_{\vect{x}\in target} \textrm{NLL}(\vect{x}) \\
      &- \frac{1}{N_{outlier}} \sum_{\vect{x}\in outlier}  \textrm{NLL}(\vect{x}) \cdot I[\textrm{NLL}(\vect{x})<c], 
  \label{new_loss}
  \end{split}
\end{equation}
where $N_{target}$ is the number of the target data in each batch and $N_{outlier}$ is the number of the 
outlier data in each batch that satisfies the condition in $I[\cdot]$. Threshold $c$ is decided 
so that the NLL of the outlier data does not go to $+\infty$. 
The difference between (\ref{kiri_loss}) and (\ref{new_loss}) is that the loss function in (\ref{kiri_loss}) 
uses out-of-distribution data that has completely different data structures from the in-distribution data, 
while the loss function in (\ref{new_loss}) uses the outlier data of the same machine type as the target data.
In (\ref{new_loss}), the NLL of the target data and 
the NLL of the outlier data that satisfies the condition in $I[\cdot]$ have the same impact on the loss.
In unsupervised anomaly detection tasks, lower NLL of the target data can lead to better detection performances.
Therefore, we modified (\ref{new_loss}) so that the model can prioritize decreasing the NLL of the target data, as
\begin{equation}
  \begin{split}
  L = &\frac{1}{N_{target}} \sum_{\vect{x}\in target} \textrm{NLL}(\vect{x}) \\
      & - k\cdot \frac{1}{N_{outlier}} \sum_{\vect{x}\in outlier}  \textrm{NLL}(\vect{x})\\ 
      & \cdot I[\textrm{NLL}(\vect{x})<c \land \textrm{NLL}(\vect{x})<\textrm {max}(\textrm{NLL}(\vect{x}_{t}))], 
  \label{mod_loss}
  \end{split}
\end{equation}
where $0<k<1$ enhances decreasing the NLL of the target data.
The second condition in $I[\cdot]$ is for removing the penalty if the NLL of the 
outlier data is larger than the maximum NLL of the target data $\vect{x}_{t}$ in each batch. 
In this case, the distinction between the target data and the outlier data can be completely made
by the NLL, and the model should only focus on decreasing the NLL of the target data.

\section{Experiments}
\label{sec:typestyle}
\subsection{Experimental conditions}
\label{ssec:condition}
We conducted experiments using DCASE 2020 Challenge Task2 development dataset. Each data is a 10-second single-channel 
16-kHz recording from one of six machine types (ToyCar, 
ToyConveyor, fan, pump, slider, valve). It has a machine ID to distinguish three to four machines
in each machine type.

To evaluate the anomaly detection performance of our proposed method, we used Glow and MAF as NF models. 
Glow and MAF are often chosen as a NF model 
in out-of-distribution detection and anomaly detection tasks \cite{Haunschmid2020}.
VAE and VIDNN \cite{Suefusa2020} were used as conventional unsupervised approaches.
MobileNetV2 \cite{Sandler2018} was used as the classifier in the self-supervised classification-based 
approach.

For the input, the frames of the log-Mel spectrograms were computed
with a length of 1024, hop size of 512, and 128 Mel bins. At least 313 frames were generated for 
each recording, and several frames were successively concatenated to make each input.
Input details and the model architectures are described as follows.

\noindent
\textbf{VAE and VIDNN. } 
Five frames were used for each input, with four overlapping frames.
We trained a model for each machine ID.
The model had ten linear layers with 128 dimensions, except for the fifth 
layer, which had eight dimensions.
The model was trained for 100 epochs using the Adam optimizer 
\cite{Kingma2014} with a learning rate of $10^{-3}$.


\noindent
\textbf{MobileNetV2. }
64 frames were used with 48 overlappings. We trained a model to 
classify machine IDs for each machine type. The width multiplier parameter
was set to 0.5. The model was trained for 20 epochs with the Adam optimizer and a learning rate of $10^{-3}$.

\noindent
\textbf{Glow.}
32 frames were used with 28 overlappings. We trained a model for each machine ID.
The model had three blocks with 12 flow steps in each block, and each flow step had 
three CNN layers with 128 hidden layers. 
The model was trained for 100 epochs using the 
AdaMax optimizer \cite{Kingma2014} with a learning rate of $5\times 10^{-4}$ and the batch size of 64. 
$k$ in  (\ref{mod_loss}) was set to $0.5$ for all machines so that the first term in (\ref{mod_loss}) could be 
lower while the second term was not ignored. 
For $c$ in (\ref{mod_loss}), we first used all data from a machine type to train the model 
for several epochs with the NLL as the loss, and then decided $c$ by the NLL in the last epoch.
The detection performance was not severely affected by the number of epochs in the first step.
This operation was performed for each machine type, and different values were set as shown 
in Table \ref{c_list}. 

\noindent
\textbf{MAF.}
Four frames were used with no overlapping. We trained a model for each machine ID.
The model had four MADE-blocks with 512 units for each block. The model was trained 
for 100 epochs using the Adam optimizer with a learning rate of $10^{-4}$ and the batch size of 64.
We set $k$ to $0.5$ for all machines, and $c$ to the values in Table \ref{c_list}.

\subsection{Results and Discussions}
\label{ssec:result}
We evaluated the detection performance on the area under the receiver operating 
characteristic curve (AUC) and partial AUC (pAUC) with $p = 0.1$. 
For NF models, we first used the conventional approach where the NLL was the loss, 
with additive/affine coupling layers and transformations (Glow add., Glow aff., MAF add., MAF aff.).
We then used our proposed methods in (\ref{new_loss}) and (\ref{mod_loss}).
We trained a model for each machine ID, where the data from that machine ID is the target data and the data from other machine IDs with 
the same machine type is the outlier data.
In Table 2, the average AUC and pAUC of each machine type for 
each model are listed. Our proposed methods outperformed all the conventional methods except for ToyConveyor.
This result indicates that our method improves the detection performance by using the outlier data 
which can have similar data structures to the target data.
For ToyConveyor, we found that clear distinctions between different IDs in ToyConveyor were made by the NLL.
For example, in ToyConveyor ID 0, the NLL using Glow with 
affine coupling layers and the loss in (\ref{mod_loss}) converged to $5.59$ for the target data and $5.83$ for the outlier data. 
We consider different machine IDs in ToyConveyor had completely different data structures, and this made 
our proposed methods ineffective for ToyConveyor
\footnote{In \cite{koizumi2020_2}, we argued that the self-supervised classification-based approach shows low AUCs on ToyConveyor  
because machine IDs in ToyConveyor have very similar data structures, and the classifier cannot classify these IDs.
However, the results in this paper show that the AUCs are low although these IDs can be distinguished easily.
These results indicate that the cause for the low scores of the self-supervised classification-based
approach is not that the machine IDs in ToyConveyor have very similar data structures.}.
The loss function in (\ref{mod_loss}) shows almost the same performance as in (\ref{new_loss}).
However, the NLL of the target data became lower with (\ref{mod_loss}) than (\ref{new_loss}).
For example, in ToyCar machine ID 0, the NLL of the target data using MAF converged to $-768$ with (\ref{new_loss}) 
and to $-782$ with (\ref{mod_loss}).
As described in \cite{Papamakarios2018}, a lower NLL means the model can obtain accurate densities of the data, 
which can improve the detection performance of the unsupervised approach.
Therefore, we can expect the loss function in (\ref{mod_loss}) to show better results than (\ref{new_loss}) for other datasets.
We also found that MAF with affine transformations could output infinities with the proposed loss function in (\ref{new_loss}) and (\ref{mod_loss}).
As previously pointed out \cite{Behrmann2020}, affine transformations have a numerical stability problem.
The affine transformations slightly improved the detection performance (Table 2),
while it simultaneously destabilized the model and required a higher computational cost. 

\begin{table}[t]
  \begin{center}
  \caption{Hyperparameter value $c$ for each machine type.}
  \begin{tabular}{lccccccc}
  \bhline{1.5pt}
    & Glow & MAF \\ \hline
  ToyCar  & 5.75 & --750   \\ 
  ToyConveyor  & 5.70 & --750   \\ 
  fan  & 5.69 & --750   \\ 
  pump  & 5.70 & --750   \\ 
  slider  & 5.63 & --750   \\ 
  valve  & 5.53 & --800   \\ \bhline{1.5pt}
  \end{tabular}
  \label{c_list}
  \end{center}
\end{table}

  \begin{table*}[t]
    \begin{center}
    \caption{Anomaly detection results for different machine types. ``total'' denotes the average scores for all 23 machine IDs.}
    \begin{tabular}{lcc|cc|cc|cc|cc|cc}
    \bhline{1.5pt}
    Model       & \multicolumn{2}{c}{VAE}   & \multicolumn{2}{c}{VIDNN} & \multicolumn{2}{c}{MAF add.} &\multicolumn{2}{c}{MAF aff.}& \multicolumn{2}{c}{Glow add.} & \multicolumn{2}{c}{Glow aff.} \\ \hline
    Metric      &  AUC       &  pAUC        & AUC             & pAUC    & AUC      &  pAUC              & AUC         & pAUC           & AUC          & pAUC            & AUC           & pAUC         \\ \hline
    ToyCar      &  83.3      &  71.3        & 81.2            &  73.1   & 79.0     &  63.2              & 79.5        & 63.9           & 75.0         & 63.6            & 75.9          & 65.1        \\ 
    ToyConveyor &  74.1      &  \bf{60.8}   & \bf{74.7}       &  58.9   & 72.3     &  58.6              & 73.3        & 60.2           & 71.6         & 59.0            & 70.7          & 58.7        \\ 
    fan         &  72.1      &  59.0        & 72.8            &  58.9   & 68.9     &  58.5              & 69.5        & 57.9           & 70.8         & 59.7            & 71.5          & 59.5         \\ 
    pump        &  76.2      &  65.6        & 76.6            &  66.9   & 76.9     &  66.2              & 78.3        & 68.1           & 75.3         & 65.8            & 76.3          & 65.6        \\ 
    slider      &  84.2      &  63.2        & 84.7            &  61.7   & 90.1     &  71.3              & 90.3        & 72.4           & 94.2         & 83.0            & 93.9          & 81.5      \\ 
    valve       &  70.2      &  51.4        & 83.0            &  60.5   & 74.6     &  54.0              & 77.4        & 55.3           & 84.9         & 63.8            & 86.1          & 67.8        \\ \bhline{1.5pt}
    total       &  76.8      &  61.9        & 79.0            &  63.5   & 77.2     &  62.1              & 78.2        & 63.1           & 78.9         & 66.1            & 79.4          & 66.7    \\ \bhline{1.5pt}
    \end{tabular}\\
    \label{result_1}
    \end{center}
    \end{table*}
    \begin{table*}[t]
      \begin{center}
    \begin{tabular}{lcc|cc|cc|cc|cc|cc}
    \bhline{1.5pt}
    \multicolumn{1}{l}{Loss}& \multicolumn{6}{c}{Ours (eq.(\ref{new_loss}))}&\multicolumn{6}{c}{Ours (eq.(\ref{mod_loss}))} \\ \hline
    \multicolumn{1}{l}{Model}& \multicolumn{2}{c}{MAF add.}& \multicolumn{2}{c}{Glow add.} & \multicolumn{2}{c}{Glow aff.}  &\multicolumn{2}{c}{MAF add.}& \multicolumn{2}{c}{Glow add.} & \multicolumn{2}{c}{Glow aff.}\\ \hline
    Metric     & AUC         & pAUC             & AUC          & pAUC              & AUC           & pAUC               &  AUC         & pAUC           & AUC          & pAUC              & AUC           & pAUC \\ \hline
    ToyCar     & 91.1        & 80.9             & 91.4         & 83.1              & 92.2          & 84.1               & 91.0         & 81.5           & 91.1         & 81.5              & \bf{92.3}     & \bf{85.0}       \\ 
    ToyConveyor& 72.3        & 59.8             & 72.3         & 59.4              & 71.5          & 59.0               & 72.5         & 59.1           & 71.7         & 59.2              & 71.2          & 58.8       \\ 
    fan        & 74.2        & 65.9             & 74.6         & 65.9              & \bf{74.9}     & 65.3               & 74.7         & \bf{66.4}      & 74.7         & 65.9              & 74.8          & 65.1       \\ 
    pump       & 82.4        & 72.0             & 82.2         & 71.3              & \bf{83.4}     & \bf{73.8}          & 82.1         & 71.7           & 82.5         & 72.3              & 83.1          & 72.8       \\ 
    slider     & 91.0        & 74.2             & 94.8         & 83.5              & 94.6          & 82.8               & 90.3         & 75.1           & \bf{95.4}    & \bf{83.9}         & 94.8          & 82.9       \\ 
    valve      & 83.1        & 62.8             & 90.1         & 70.9              & \bf{91.4}     & \bf{75.0}          & 81.3         & 61.1           & 89.7         & 70.4              & 90.5          & 74.1       \\ \bhline{1.5pt}
    total      & 82.8        & 69.7             & 84.8         & 72.9              & \bf{85.2}     & \bf{73.9}          & 82.4         & 69.6           & 84.7         & 72.7              & 85.0          & 73.7       \\ \bhline{1.5pt}
          \end{tabular}
      \label{result_1}
      \end{center}
      \end{table*}

\begin{table}[]
  \begin{center}
  \caption{Minimum AUC for the self-supervised cla\-s\-s\-i\-f\-i\-c\-a\-tion-based approach (MobileNetV2), the unsupervised approach (Glow aff.), and our method with Eq. (\ref{mod_loss}) (Glow aff. (\ref{mod_loss})).}
  \begin{tabular}{p{1.7cm}ccc}
  \bhline{1.5pt}
              & MobileNetV2 & Glow aff.  & Glow aff. (\ref{mod_loss}) \\ \hline
  ToyCar      & 55.7        &  64.2      &         \bf{80.1}   \\ 
  ToyConveyor & 48.7        &  60.9      &         \bf{61.0}   \\ 
  fan         & \bf{50.4}   &  49.6      &         49.6   \\ 
  pump        & 52.9        &  63.2      &         \bf{65.7}   \\ 
  slider      & 82.8        &  84.5      &         \bf{87.8}   \\ 
  valve       & 67.9        &  67.2      &         \bf{77.7}   \\ \bhline{1.5pt}
  \end{tabular}
  \label{MobileV2}
  \end{center}
\end{table}

\begin{table}[]
  \begin{center}
  \caption{AUC of pump ID 0, where the target data is pump ID 0 and the outlier data is from each machine type.}
  \begin{tabular}{llcc}
  \bhline{1.5pt}
  Method& Outlier data & AUC\\ \hline
  \begin{tabular}[c]{@{}l@{}l@{}}Ours (Glow aff. (\ref{mod_loss}) \\using the same machine \\type for the outlier data) \end{tabular}& pump (except ID 0)&\bf{74.2}\\ \hline
  \multirow{5}{*}{\begin{tabular}[c]{@{}l@{}l@{}}Glow aff. (\ref{mod_loss}) using a \\different machine type\\for the outlier data \end{tabular}}&ToyCar  & 70.0   \\ 
  &ToyConveyor  & 70.5   \\ 
  &fan  & 71.1  \\ 
  &slider  & 69.4   \\ 
  &valve  & 69.7   \\ \hline
  Minimizing NLL& - & 69.6  \\ \bhline{1.5pt}
  \end{tabular}
  \label{similar}
  \end{center}
\end{table}

In Table \ref{MobileV2}, we present the results of a machine ID with minimum AUC for each machine type.
The minimum AUC of the self-supervised classification-based approach was significantly lower than the 
unsupervised approach for three out of the six machine types, while that of our method was equal to or significantly 
higher than the unsupervised approach.
These results indicate that our proposed method not only outperforms the unsupervised approach but also
shows more stable detection performances than the self-supervised classification-based approach, and therefore can be 
suitable for practical applications.

We also show experimentally that our method does not improve the detection performance if the outlier data has different
data structures from the target data. We used pump machine ID 0 as the target data and made six models in which
the outlier data was from each machine type. The model was trained with the loss in  (\ref{mod_loss}).
As shown in Table \ref{similar}, the detection performance did not improve if the outlier data was from a machine type
other than pump. These results indicate that our method does not improve the detection performance if the outlier data has
different structures from the target data.

\section{Conclusion}
We proposed flow-based methods for anomalous sound detection that outperform
unsupervised approaches while maintaining greater stability than the self-supervised classif\-i\-c\-a\-t\-ion-based approach.
Experimental results demonstrated that our methods improve the detection performance
by using the outlier data from the same machine type as the target data.
Our future work will include the development of more efficient methods to decide the hyperparameters.
\clearpage
\bibliographystyle{IEEEbib}
\bibliography{refs}

\end{document}